\documentclass[12pt]{iopart}
\usepackage{graphicx}
\begin{document}
\title[Percolation on four-dimensional simple hypercubic lattices]{Site and bond percolation on four-dimensional simple hypercubic lattices with extended neighborhoods}

\author{Pengyu Zhao$^{1}$\footnote[1]{These two authors contributed equally to this paper}
Jinhong Yan$^{1}\footnotemark[1]$,
Zhipeng Xun$^{1}$\footnote{zpxun@cumt.edu.cn},
Dapeng Hao$^1$,
Robert M. Ziff$^2$}
\address{$^1$ School of Material Sciences and Physics, China University of Mining and Technology, Xuzhou 221116, China}
\address{$^2$ Center for the Study of Complex System and Department of Chemical Engineering, University of Michigan, Ann Arbor, Michigan 48109-2800, USA}

\begin{abstract}
The asymptotic behavior of the percolation threshold $p_c$ and its dependence upon coordination number $z$ is investigated for both site and bond percolation on four-dimensional lattices with compact extended neighborhoods. Simple hypercubic lattices with neighborhoods up to 9th nearest neighbors are studied to high precision by means of Monte-Carlo simulations based upon a single-cluster growth algorithm. For site percolation, an asymptotic analysis confirms the predicted behavior $zp_c \sim 16 \eta_c = 2.086$ for large $z$, and finite-size corrections are accounted for by forms $p_c \sim 16 \eta_c/(z+b)$ and $p_c \sim 1- \exp(-16 \eta_c/z)$ where $\eta_c \approx 0.1304$ is the continuum percolation threshold of four-dimensional hyperspheres. For bond percolation, the finite-$z$ correction is found to be consistent with the prediction of Frei and Perkins, $zp_{c} - 1 \sim a_{1} (\ln z)/z$, although the behavior $zp_{c} - 1 \sim a_1 z^{-3/4}$ cannot be ruled out. 
\end{abstract}


\maketitle

\section{Introduction}
Percolation on lattices with extended neighborhoods, which goes back to the ``equivalent neighbor model" of Dalton, Domb and Sykes in 1964 \cite{DaltonDombSykes64,DombDalton1966,Domb72}, has been a question of long-standing interest in the percolation field, both because of its theoretical significance and practical applications \cite{Kleinberg2000,SanderWarrenSokolov2003,Ziff2021,KozaKondratSuszcaynski2014,KozaPola2016}. Following the work of Dalton, Domb and Sykes, many researchers have performed extensive investigations, such as long-range site percolation on compact regions in a diamond shape on a square lattice \cite{GoukerFamily83}, site and bond percolation on body-centered cubic lattices with nearest and next-nearest neighbors \cite{JerauldScrivenDavis1984}, site percolation on face-centered cubic lattices up to fourth nearest neighbors \cite{GawronCieplak91}, site percolation on all eleven of the Archimedian lattices with long-range connections \cite{dIribarneRasigniRasigni95,dIribarneRasigniRasigni99,dIribarneRasigniRasigni99b}. The idea of ``complex neighborhoods" where various combinations of neighborhoods, not necessarily compact, was introduced by Malarz and Galam \cite{MalarzGalam05}, and has been followed up by many subsequent investigations in two, three, and four dimensions for both site percolation \cite{MalarzGalam05,MajewskiMalarz2007,KurzawskiMalarz2012,Malarz2015,KotwicaGronekMalarz19,Malarz2020,XunHaoZiff2021} and bond percolation \cite{OuyangDengBlote2018,DengOuyangBlote2019,XunZiff2020,XunZiff2020b,XuWangHuDeng2021}.  In the mathematics field, there has also been work on this model, under the name of ``range-$R$" model.  Penrose \cite{Penrose93} studied both site and bond percolation, including asymptotic behavior for bond percolation, and more recently Frei and Perkins \cite{FreiPerkins2016} and others \cite{Hong21} have studied the finite-size corrections.

Since the early days of the study of percolation, researchers have focused on exploring the correlations between percolation thresholds $p_c$ and properties of the lattice, especially the coordination number $z$ which gives the number of nearest neighbors to a vertex, since $z$ (along with the dimensionality) seems to be one of the main determinants of the threshold.  
In the context of extended-range percolation, earlier investigations \cite{GalamMauger1996,vanderMarck1998,KurzawskiMalarz2012,XunZiff2020,XunZiff2020b} demonstrated that the percolation thresholds for lattices with extended neighborhoods can be fitted well by a simple power given by $p_c \sim z^{-a}$ or $p_c \sim (z-1)^{-a}$. It has been argued \cite{Domb72,dIribarneRasigniRasigni99b,KozaKondratSuszcaynski2014,KozaPola2016} that for longer range site percolation, the asymptotic behavior for large $z$ could be related to the continuum percolation threshold $\eta_c$ for objects of the same shape as the neighborhood. In general $d$, these arguments imply \cite{XunHaoZiff2021}
\begin{equation}
    zp_c \sim 2^d \eta_c,
\label{eq:asysite}
\end{equation}
where $d$ is the dimension of the system. Equation (\ref{eq:asysite}) holds for systems with compact neighborhoods, and is independent of the type of lattice.  For finite $z$, one expects asymptotic corrections of the form
\begin{equation}
    p_c \sim \frac{c}{z+b},
\label{eq:finitesite}
\end{equation}
where $c=2^d \eta_c$ and $b$ an empirical constant.  Another proposed form (without any additional parameters) is \cite{XunHaoZiff2021}
\begin{equation}
    p_c \sim 1- \exp\left(-\frac{2^d \eta_c}{z}\right).
\label{eq:finitesiteb}
\end{equation}
For bond percolation, the asymptotic behavior of $p_c$ tends to the Bethe-lattice behavior \cite{Penrose93}
\begin{equation}
    p_c \sim \frac{1}{z-1},
\label{asybond}
\end{equation}
for $z \to \infty$, because for large $z$ and the corresponding low $p$, the chance of hitting the same site twice is vanishingly small and the system behaves basically like a tree. The finite-$z$ correction in four dimensions was recently predicted to be \cite{FreiPerkins2016,Hong21} 
\begin{equation}
    zp_c - 1 \sim a_1 (\ln z)/z,
\label{eq:zpclnz}
\end{equation}
This is in contrast to the behavior for $d = 2$ and 3, where it has been shown that 
\begin{equation}
zp_c - 1 \sim a_1 z^{-x},
\label{eq:zpcx}
\end{equation}
where $x = (d-1)/d$ for these two dimensions \cite{FreiPerkins2016}. In two and three dimensions, both the asymptotic behavior and the finite-$z$ correction for site and bond percolation have been confirmed by extensive simulation work \cite{dIribarneRasigniRasigni99b,OuyangDengBlote2018,DengOuyangBlote2019,XunZiff2020b,XunHaoZiff2021,XunHaoZiff2021b}. However, more work is needed to confirm the asymptotic behavior for $d \ge 4$, and to understand the difference of the correction exponents in different dimensions. 

In this paper, we focus on four-dimensional simple hypercubic (\textsc{sc(4)}) lattices with extended neighborhoods. Both site and bond percolation on \textsc{sc(4)} lattices with up to 9th nearest neighbors are investigated by employing Monte-Carlo simulation, and we find site and bond thresholds of these lattices with high precision. For site percolation, both $z$ versus $1/p_c$ and $z$ versus $-1/\ln(1-p_c)$ lead to the predicted asymptotic value of $zp_c \sim 16 \eta_c = 2.086$. For bond percolation, data fitting shows that the thresholds are consistent the behavior in Eq.\ (\ref{eq:zpclnz}) above. However, the behavior of Eq.\ (\ref{eq:zpcx}) with $x = 3/4$ cannot be ruled out.

The remainder of the paper is organized as follows. The simulation results and discussions are given in Sec. \ref{sec:simulation}, including some simulation details (\ref{sec:detail}), results for site percolation (\ref{sec:site}) and bond percolation (\ref{sec:bond}). In Sec. \ref{sec:conclusion}, we present our conclusions. 

\section{Simulation results and discussions}
\label{sec:simulation}
\subsection{Simulation details and basic theory}
\label{sec:detail}
We use the notation \textsc{sc(4)}-$a, b, ...$ to indicate a four-dimensional simple hypercubic lattice with the $a$th nearest neighbors, $b$th nearest neighbors, etc. In the Monte-Carlo simulations, by using a single-cluster growth algorithm (see Refs.\ \cite{LorenzZiff1998,XunZiff2020,XunZiff2020b} for more details), a site on the lattice of size $L\times L \times L \times L$ with $L=128$ under periodic boundary conditions is chosen as the seed, and an individual cluster is grown from that seeded site. We grow many samples of individual clusters for each lattice, which are $5 \times 10^8$ for \textsc{sc(4)}-3, \textsc{sc(4)}-1,3, \textsc{sc(4)}-2,3, \textsc{sc(4)}-1,2, \textsc{sc(4)}-1,2,3, and \textsc{sc(4)}-1,2,3,4 lattices, and $10^8$ for other five lattices. Clusters with different sizes are distributed in bins of range of $(2^n, 2^{n+1}-1)$ for $n=0,1,2,\ldots$. An upper size of the cluster, which is called the upper size cutoff, needs to be set to halt the growth of clusters larger than that size, to avoid wrapping around the boundaries. For clusters still growing when they reach an upper size cutoff, we count them in the last bin. Here for site percolation, we set the upper size cutoff to be $2^{15}$ occupied sites for the \textsc{sc(4)}-1,2 and \textsc{sc(4)}-1,2,3 lattices, $2^{13}$ for the \textsc{sc(4)}-1,...,9 lattice, and $2^{14}$ for other lattices. For bond percolation, we set the upper size cutoff to be $2^{16}$ occupied (wetted) sites for all lattices.

If one defines $n_{s}(p)$ as the number of clusters (per site) containing $s$ occupied sites, as a function of the site or bond occupation probability $p$, then in the scaling limit, in which $s$ is large and $(p-p_{c})$ is small such that $(p-p_{c})s^\sigma$ is constant, $n_{s}(p)$ behaves as
\begin{equation}
    n_{s}(p) \sim A_0 s^{-\tau} f[B_0 (p-p_{c}) s^ \sigma],
    \label{eq:nsp}
\end{equation}
where $\tau$, $\sigma$, and $f(x)$ are universal, while $A_0$ and $B_0$ are lattice-dependent metric factors. When $p = p_c$ and for finite $s$, there are corrections to Eq.\ (\ref{eq:nsp})
\begin{equation}
    n_{s}(p_{c}) \sim A_0 s^{-\tau} (1+C_0 s^{-\Omega}+\dots),
    \label{eq:finite}
\end{equation}
where $\Omega$ is another universal exponent. Then the probability that a point belongs to a cluster of size greater than or equal to $s$ is given by $P_{\ge s} = \sum_{s'=s}^\infty s' n_{s'}$, and it follows from Eqs.\ (\ref{eq:nsp}) and (\ref{eq:finite}) that \cite{XunZiff2020,XunZiff2020b} 
\begin{equation}
    s^{\tau - 2} P_{\geq s} \sim A_1 [1 + B_1 (p-p_{c}) s^ \sigma+C_1 s^{-\Omega} + \dots],
    \label{eq:nsp2}
\end{equation}
where $A_1$, $B_1$ and $C_1$ are non-universal constants. Equation (\ref{eq:nsp2}) provides two methods to determine the percolation threshold $p_c$, and we will show them in detail in the next subsection, combined with our simulation results.

Due to the universality of $\tau$, $\Omega$ and $\sigma$, these exponents depend only on the system dimension, not the type of the lattice. We choose the central values of the estimates $\tau = 2.3135(5)$\cite{XunZiff2020}, $\Omega = 0.40(3)$\cite{XunZiff2020}, and $\sigma = 0.4742$\cite{Gracey2015,BorinskyGraceyKompanietsSchnetz21}, which are relatively accurate and acceptable, in the numerical simulations. The number of clusters greater than or equal to size $s$ could be found based on the data from our simulation, and the quantity $s^{\tau-2}P_{\geq s}$ could be easily calculated.

\subsection{Results of site percolation}
\label{sec:site}
\begin{figure}[htbp] 
\centering
\includegraphics[width=4in]{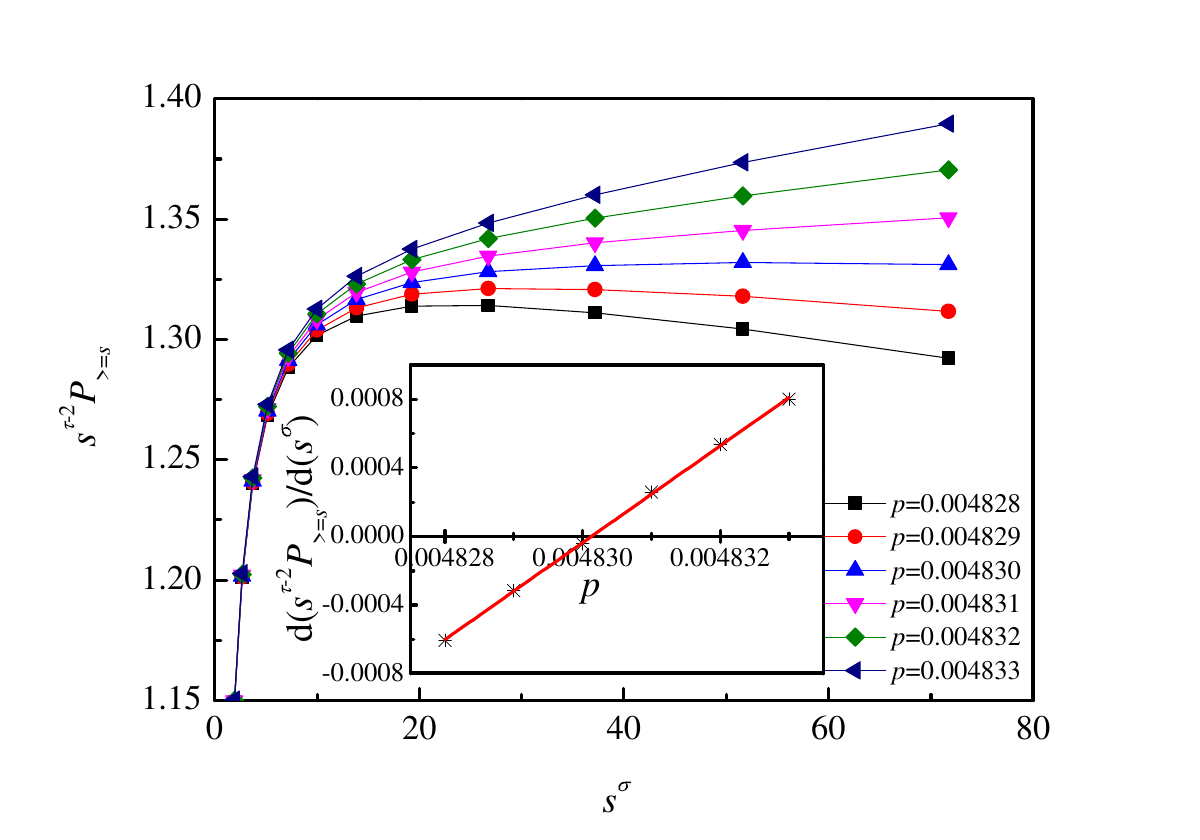}
\caption{Plot of $s^{\tau-2}P_{\geq s}$ versus $s^{\sigma}$ with $\tau = 2.3135$ and $\sigma = 0.4742$ for the site percolation of the \textsc{sc(4)}-1,...,9 lattice under different values of $p$. The inset indicates the slope of the linear portions of the curves shown in the main figure as a function of $p$, and the central value of $p_{c} = 0.0048301$ can be calculated from the $p$ intercept.}
\label{fig:sc4-1-9-sigma-site}
\end{figure}

From the behavior of Eq.\ (\ref{eq:nsp2}), we can determine if we are above, near, or below the percolation threshold. For large $s$ where the finite-size effect term $s^{-\Omega}$ can be ignored, Eq.\ (\ref{eq:nsp2}) becomes
\begin{equation}
    s^{\tau - 2} P_{\geq s} \sim A_1 [1 + B_1 (p-p_{c})s^\sigma].
\end{equation}
This implies that $s^{\tau-2}P_{\geq s}$ will convergence to a constant value at $p_{c}$, while it deviates linearly from that constant value when $p$ is away from $p_{c}$, when plotted as a function of $s^\sigma$. Figure \ref{fig:sc4-1-9-sigma-site} shows the relation of $s^{\tau-2}P_{\geq s}$ versus $s^{\sigma}$ for the site percolation of the \textsc{sc(4)}-1,...,9 lattice under probabilities $p = 0.004828$, $0.004829$, $0.004830$, $0.004831$, $0.004832$, and $0.004833$. For small clusters, $s^{\tau-2}P_{\geq s}$ shows a steep rise due to the finite-size effect, while for large clusters, $s^{\tau-2}P_{\geq s}$ shows a linear region. As $p$ tends to $p_c$, the linear part of $s^{\tau-2}P_{\geq s}$ become more nearly horizontal. Based upon these properties of the linear portions in Fig.\ \ref{fig:sc4-1-9-sigma-site}, the central value of $p_c$ can be determined from the slope:
\begin{equation}
\frac{\mathrm{d} (s^{\tau-2}P_{\geq s})}{\mathrm{d} (s^{\sigma})} \sim B_1(p-p_{c}),
\label{eq:Ps}
\end{equation}
As shown in the inset of Fig.\ \ref{fig:sc4-1-9-sigma-site}, $p_{c} = 0.0048301$ can be deduced from the $p$ intercept of the plot of the above derivative versus $p$.

\begin{figure}[htbp] 
\centering
\includegraphics[width=4in]{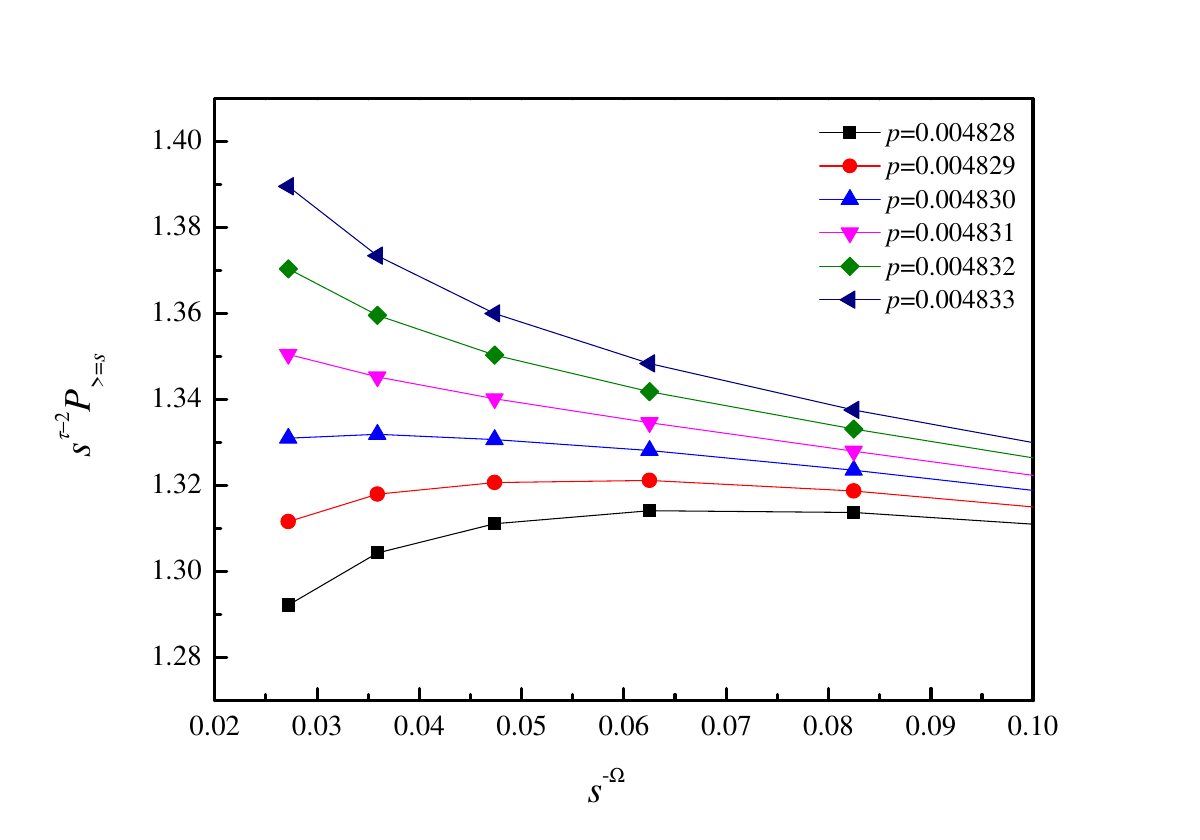}
\caption{Plot of $s^{\tau-2}P_{\geq s}$ versus $s^{-\Omega}$ with $\tau = 2.3135$ and $\Omega = 0.40$ for the site percolation of the \textsc{sc(4)}-1,...,9 lattice under different values of $p$.}
\label{fig:sc4-1-9-omega-site}
\end{figure}

If $p$ is very close to $p_c$, Eq.\ (\ref{eq:nsp2}) reduces to
\begin{equation}
    s^{\tau - 2} P_{\geq s} \sim A_1 [1 + C_1 s^{-\Omega}],
\end{equation}
implying a linear relationship between $s^{\tau-2}P_{\geq s}$ and $s^{-\Omega}$ for large $s$. In Fig.\ \ref{fig:sc4-1-9-omega-site}, we show the plot of $s^{\tau-2}P_{\geq s}$ versus $s^{-\Omega}$ for the site percolation of the \textsc{sc(4)}-1,...,9 lattice under probabilities $p = 0.004828$, $0.004829$, $0.004830$, $0.004831$, $0.004832$, and $0.004833$. It can be seen that when $p$ is away from $p_c$, the curves show an obvious deviation from linearity for large $s$, while better linear behavior emerges if $p$ is very close to $p_{c}$. Then we can conclude the range of the site percolation threshold of $0.004830 < p_{c} < 0.004831$ here.  

Based upon the two methods indicated in Figs. \ref{fig:sc4-1-9-sigma-site} and \ref{fig:sc4-1-9-omega-site}, as well as the errors for the values of $\tau = 2.3135(5)$ and $\Omega = 0.40(3)$, finally, the site percolation threshold of the \textsc{sc(4)}-1,...,9 lattice can be deduced to be $p_{c} = 0.0048301(9)$, where the number in parentheses represents the estimated error in the last digit. The figures for the other seven lattices we simulated are shown in the Supplementary Material \cite{XunZiff2020supplementary} in Figs.\ 1-14, and the corresponding site percolation thresholds are summarized in Table \ref{tab:thresholds4dsite}. 

In Table \ref{tab:thresholds4dsite}, we also show in the last column previous site percolation thresholds for the \textsc{sc(4)}-1,2 and \textsc{sc(4)}-1,2,3 lattices provided by Ref.\ \cite{KotwicaGronekMalarz19}. For these two lattices, we get substantially more precise values. For the other six lattices, it appears that our results are new.

\begin{table}[htb]
\centering
\caption{Site percolation thresholds on four-dimensional simple hypercubic lattice with extended neighborhoods up to the 9th nearest neighbors. The interaction range $R$ for the \textsc{sc}(4)-1,...,$n$ lattice is $\sqrt{n}.$}
\begin{tabular}{cccc}
\hline\hline
    lattice                       & $z$   & $p_{c}$(present)    & $p_{c}$(previous)     \\ \hline
    \textsc{sc(4)}-1,2            & 32    &  0.0617731(19)      & 0.06190(23)\cite{KotwicaGronekMalarz19}           \\
    \textsc{sc(4)}-1,2,3          & 64    &  0.0319407(13)      & 0.03190(23)\cite{KotwicaGronekMalarz19}           \\
    \textsc{sc(4)}-1,2,3,4        & 88    &  0.0231538(12)      & ---                   \\ 
    \textsc{sc(4)}-1,...,5        & 136   &  0.0147918(12)      & ---                   \\
    \textsc{sc(4)}-1,...,6        & 232   &  0.0088400(10)      & ---                   \\
    \textsc{sc(4)}-1,...,7        & 296   &  0.0070006(6)       & ---                   \\
    \textsc{sc(4)}-1,...,8        & 320   &  0.0064681(9)       & ---                   \\
    \textsc{sc(4)}-1,...,9        & 424   &  0.0048301(9)       & ---                   \\
\hline\hline
\end{tabular}
\label{tab:thresholds4dsite}
\end{table}

With regard to the asymptotic behavior between percolation thresholds $p_c$ and coordination number $z$ for site percolation on lattices with compact nearest neighborhoods in four dimensions where $\eta_c$ for hyperspheres equals $0.1304(5)$ \cite{TorquatoJiao2012}, one should expect from Eq.\ (\ref{eq:asysite}) that
\begin{equation}
    p_c = \frac{2.086(8)}{z}.
    \label{eq:pch}
\end{equation}
The extended-range nearest-neighbor regions on a lattice are not exactly spherical in shape, of course.  A limit of a more non-spherical surface is a system aligned hypercubes, where $\eta_c = 0.1201(6)$ \cite{TorquatoJiao2012}, implying $c = z p_c = 16 \eta_c = 1.922(10)$, which is not much smaller than the value above. Indeed, for some small values of $z$ considered here, the neighborhood tends to a hypercube.  However, for large $z$ where the neighborhood becomes more spherical, one would expect that $z p_c$ should approach the value 2.086.

\begin{figure}[htbp] 
\centering
\includegraphics[width=4in]{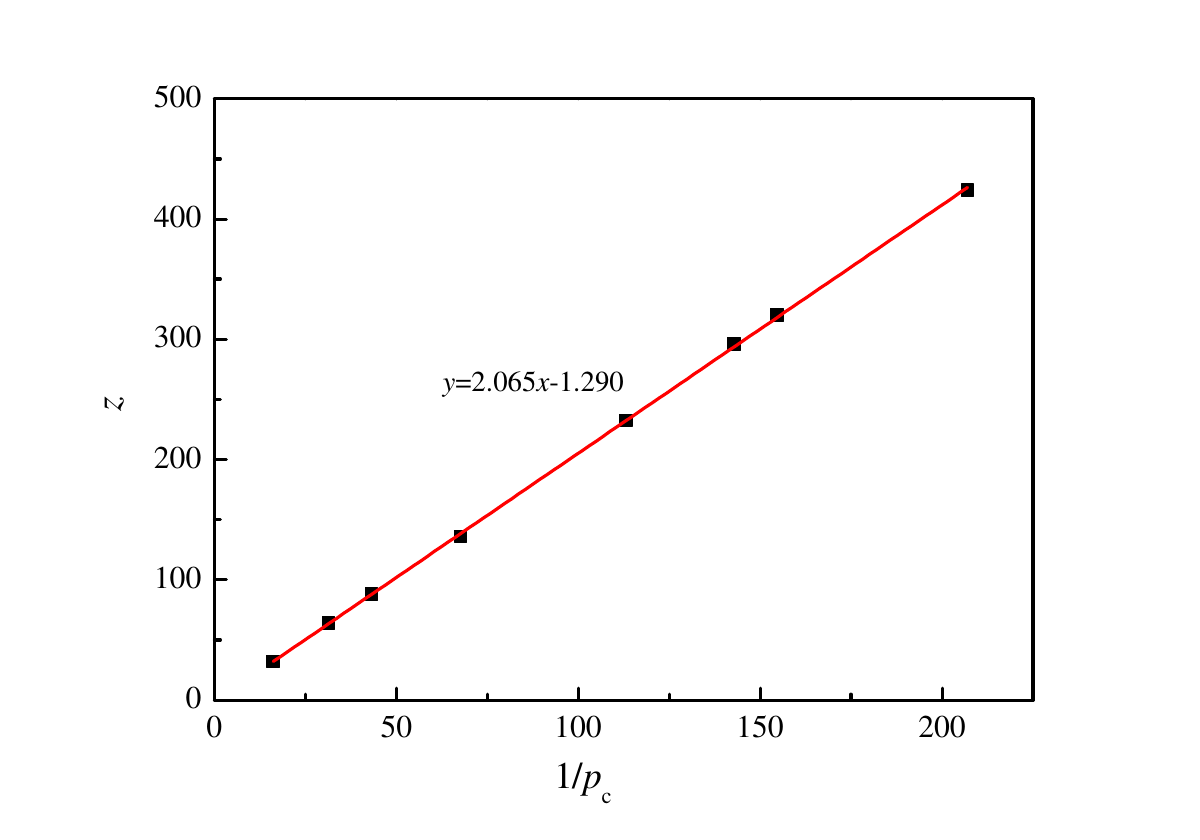}
\caption{Plot of $z$ versus $1/p_c$ for site percolation on the four-dimensional simple hypercubic lattices with compact nearest neighborhoods shown in Table \ref{tab:thresholds4dsite}. Data fitting gives the slope of $c=2.065(11)$, compared with the prediction of $zp_c \sim 16 \eta_c = 2.086(8)$, and the intercept of $ -1.290(12)$.}
\label{fig:z-vs-pc-site}
\end{figure}

\begin{figure}[htbp] 
\centering
\includegraphics[width=4in]{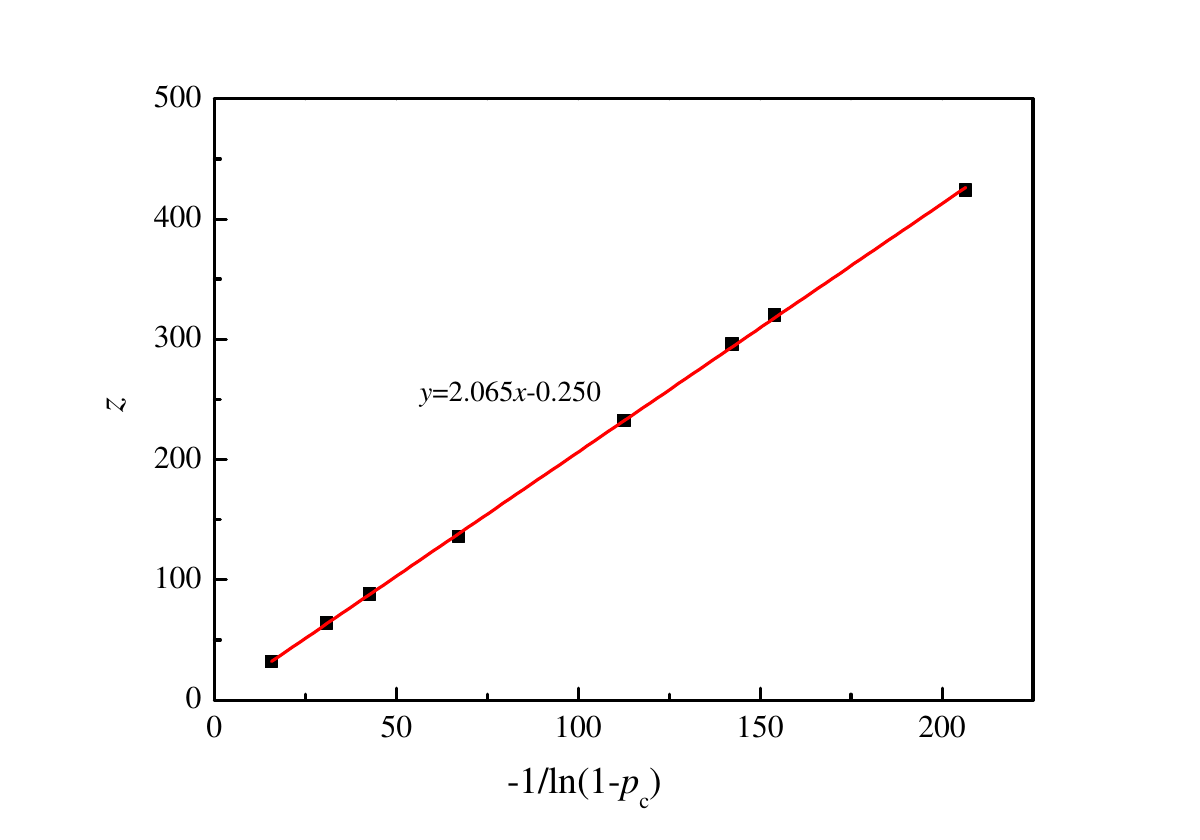}
\caption{Plot of $z$ versus $-1/\ln(1-p_c)$ for site percolation on the four-dimensional simple hypercubic lattices with compact nearest neighborhoods shown in Table \ref{tab:thresholds4dsite}. Data fitting gives the slope of $2.065(11)$, compared with the prediction of $zp_c \sim 16 \eta_c = 2.086(8)$, and the intercept of $-0.250(12)$.}
\label{fig:z-vs-lnpc-site}
\end{figure}

By plotting the relation of $z$ versus $1/p_c$ and $z$ versus $-1/\ln(1-p_c)$, as shown in Figs.\ \ref{fig:z-vs-pc-site} and \ref{fig:z-vs-lnpc-site}, respectively, we can see that data fitting of both plots lead to the asymptotic value of $2.065(11)$, which agrees quite well with the theoretically predicted value for hyperspheres above.



In contrast to the case of two and three dimensions \cite{XunHaoZiff2021b}, here we find that the plots of $z$ versus $1/p_c$ and $z$ versus $-1/\ln(1-p_c)$ have very similar behavior and give an identical slope. This is because the values of $z$ considered here are much larger than we considered in the lower dimensions, and for large $z$ and small $p_c$ the formulas are nearly identical.  If we compare the slopes and intercepts of the two plots:
\begin{eqnarray}
    z &=& \frac{c}{p_c} - b  \qquad \hbox{and}\cr
    z &=& \frac{-c'}{\ln(1-p_c)} - b' \cr
    &=& c' \left(\frac{1}{p_c} -\frac12 - \frac{p_c}{12} \ldots\right)-b'
\end{eqnarray}
we see, ignoring ${\mathcal O}(p_c)$ and higher-order terms, that the two plots are equivalent, with $c'=c$ and $b'=b-c'/2$.  In fact, our measured values $c = c' =2.065(11)$, $b = 1.290(12)$ and $b' = 0.250(12)$ agree with these formulas.  For systems of large $z$, the general formula Eq.\ (\ref{eq:finitesite}) is sufficient and for fitting the data, and the formula (\ref{eq:finitesiteb}) is not more beneficial.


\subsection{Results of bond percolation}
\label{sec:bond}
\begin{figure}[htbp] 
\centering
\includegraphics[width=4in]{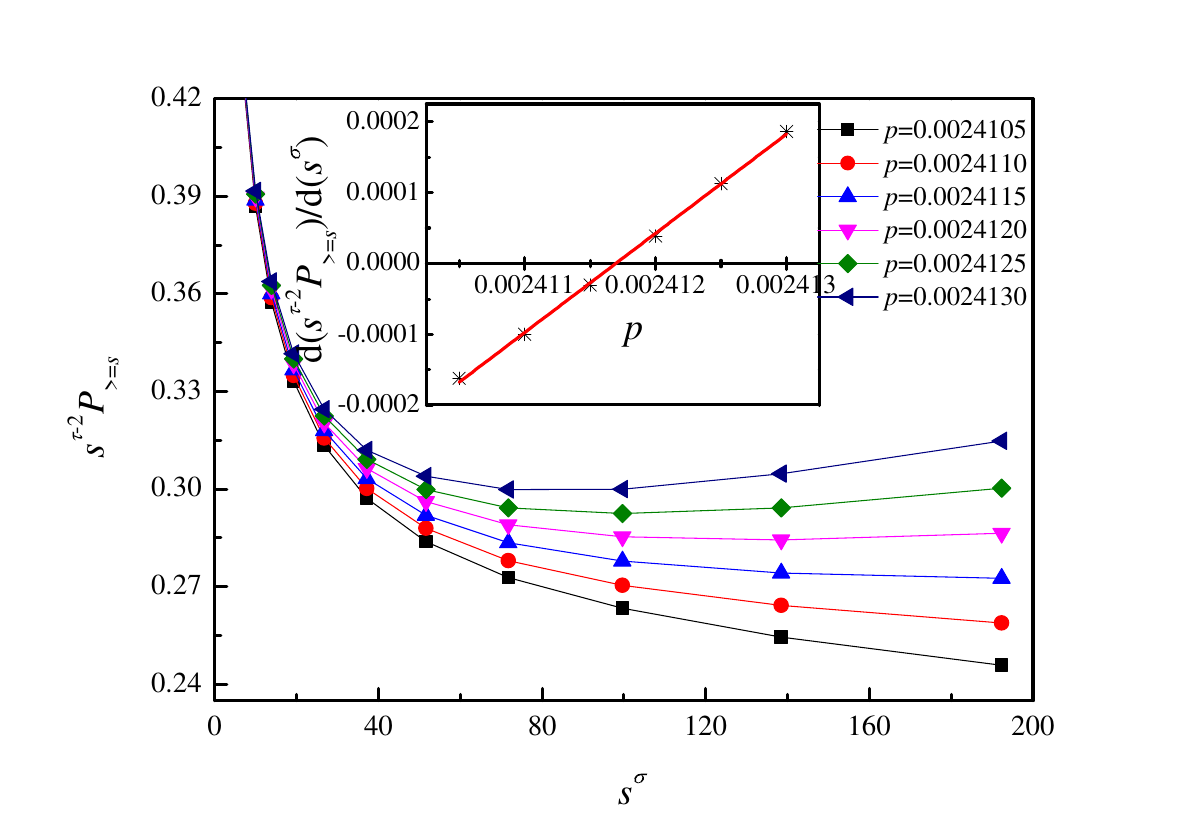}
\caption{Plot of $s^{\tau-2}P_{\geq s}$ versus $s^{\sigma}$ with $\tau = 2.3135$ and $\sigma = 0.4742$ for bond percolation of the \textsc{sc(4)}-1,...,9 lattice under different values of $p$. The inset indicates the slope of the linear portions of the curves shown in the main figure as a function of $p$, and the central value of $p_{c} = 0.0024117$ can be calculated from the $p$ intercept.}
\label{fig:sc4-1-9-sigma-bond}
\end{figure}

\begin{figure}[htbp] 
\centering
\includegraphics[width=4in]{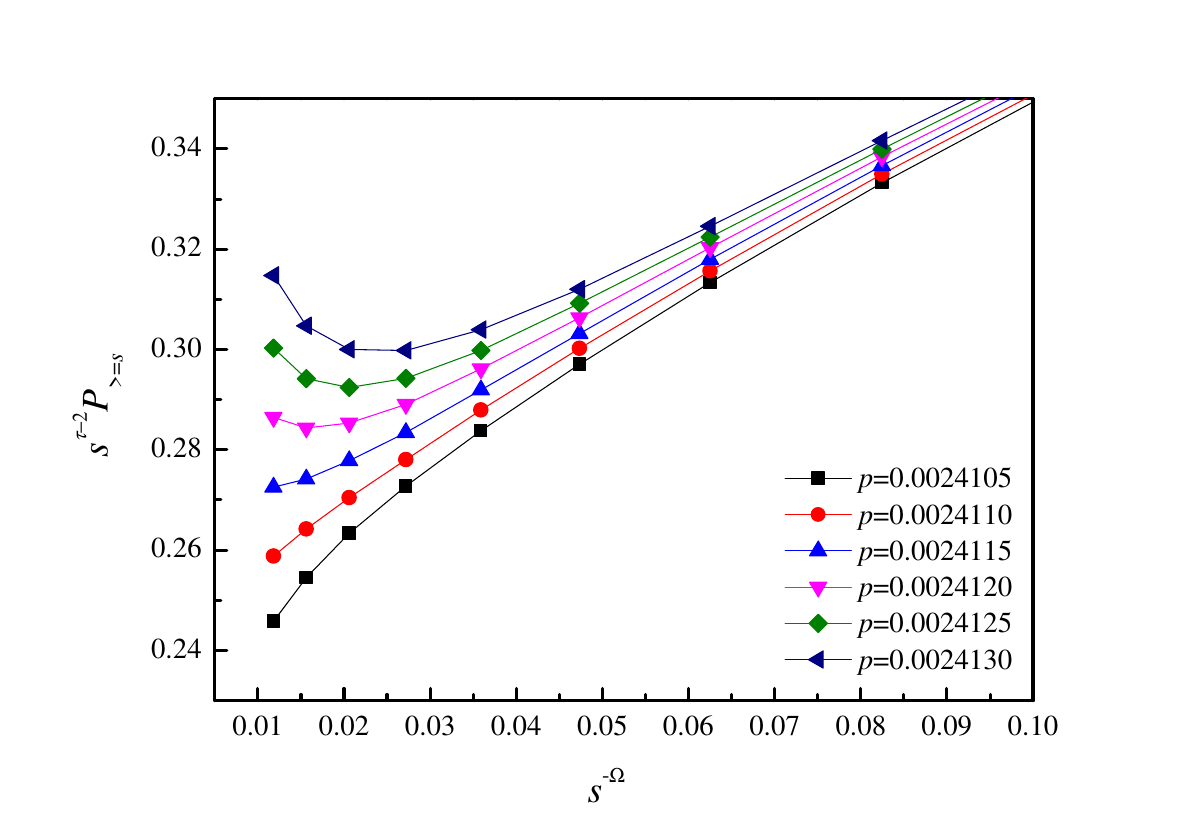}
\caption{Plot of $s^{\tau-2}P_{\geq s}$ versus $s^{-\Omega}$ with $\tau = 2.3135$ and $\Omega = 0.40$ for bond percolation of the \textsc{sc(4)}-1,...,9 lattice under different values of $p$.}
\label{fig:sc4-1-9-omega-bond}
\end{figure}

For the bond percolation simulated in this paper, Figs.\ \ref{fig:sc4-1-9-sigma-bond} and \ref{fig:sc4-1-9-omega-bond} show the plots of $s^{\tau-2}P_{\geq s}$ versus $s^{\sigma}$ and $s^{-\Omega}$, respectively, for the \textsc{sc(4)}-1,...,9 lattice under probabilities $p = 0.0024105$, $0.0024110$, $0.0024115$, $0.0024120$, $0.0024125$, and $0.0024130$. Similar to the discussion above, we conclude the bond percolation threshold of the \textsc{sc(4)}-1,...,9 lattice here to be $p_c = 0.0024117(7)$. The figures for the other 9 lattices we simulated are shown in the Supplementary Material \cite{XunZiff2020supplementary} in Figs. 15-32, and the corresponding bond percolation thresholds are summarized in Table \ref{tab:thresholds4dbond}.

\begin{table}[htb]
\centering
\caption{Bond percolation thresholds for the four-dimensional simple hypercubic lattice with extended neighborhoods up to the 9th nearest neighbors.  Previous results include the \textsc{sc(4)}-1,2 ($z=32$, $p_c = 0.035827(1)$), \textsc{bcc} = \textsc{sc(4)-4} ($z=16$), and \textsc{fcc} = \textsc{sc(4)-2} ($z=24$) lattices \cite{XunZiff2020}.}
\begin{tabular}{cccc}
\hline\hline
    lattice                       & $z$   & $p_{c}$        & $zp_c$       \\ \hline
    \textsc{sc(4)}-3              & 32    & 0.0338047(27)  & 1.08175   \\
    \textsc{sc(4)}-1,3            & 40    & 0.0271892(22)  & 1.08757   \\
    \textsc{sc(4)}-2,3            & 56    & 0.0194075(15)  & 1.08682   \\
    \textsc{sc(4)}-1,2,3          & 64    & 0.0171036(11)  & 1.09463   \\
    \textsc{sc(4)}-1,2,3,4        & 88    & 0.0122088(8)   & 1.07437   \\ 
    \textsc{sc(4)}-1,...,5        & 136   & 0.0077389(9)   & 1.05249   \\
    \textsc{sc(4)}-1,...,6        & 232   & 0.0044656(11)  & 1.03602   \\
    \textsc{sc(4)}-1,...,7        & 296   & 0.0034812(7)   & 1.03044   \\
    \textsc{sc(4)}-1,...,8        & 320   & 0.0032143(8)   & 1.02858   \\
    \textsc{sc(4)}-1,...,9        & 424   & 0.0024117(7)   & 1.02256   \\
\hline\hline
\end{tabular}
\label{tab:thresholds4dbond}
\end{table}

\begin{figure}[htbp] 
\centering
\includegraphics[width=4 in]{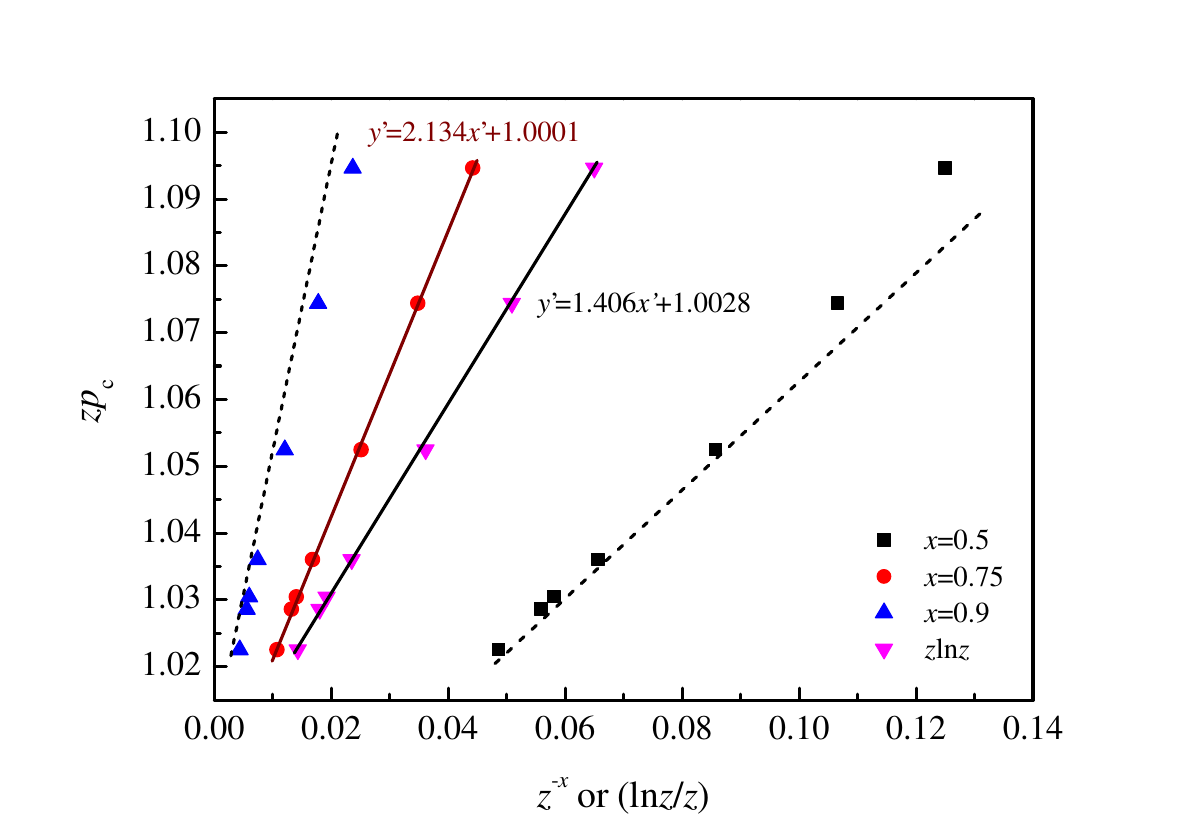}
\caption{Plot of $zp_c$ versus $z^{-x}$ with $x = 0.5$, $0.75$, and $0.9$, and versus $\ln z/z$, for bond percolation on the simple hypercubic lattices with compact nearest neighborhoods listed in Table \ref{tab:thresholds4dbond} with $z\ge 64$. Data fittings show that both $z^{-3/4}$ and $(\ln z)/z$ give good representations of the behavior of $z p_c$. The intercept of the brown line for the fit with $x=3/4$, $1.0001(5)$ is close to 1 as required, and the slope gives $a_1 = 2.134(19)$. The intercept and slope of the black line for the fit with $\ln z/z$ are $1.0028(10)$ and $1.406(14)$, respectively.}
\label{fig:zpc-vs-z-bond}
\end{figure}

Table \ref{tab:thresholds4dbond} also shows the values of $zp_c$ for each lattice. With the increase of $z$, the value of $zp_c$ decreases and tends to the asymptotic value of 1. In Fig.\ \ref{fig:zpc-vs-z-bond}, we show the relation of $zp_c$ versus $z^{-x}$ with $x = 0.5$, 0.75, and 0.9, and also versus $(\ln z)/z$. We see that the behavior is consistent with the prediction of Eq.\ (\ref{eq:zpclnz}), but we cannot really rule out the behavior of Eq.\ (\ref{eq:zpcx}) with $x = 3/4$, which follows the behavior $x = (d-1)/d$ predicted for $d = 2$ and 3.  


\section{Conclusion\textbf{}}
\label{sec:conclusion}
To summarize, in this paper, in order to further explore the correlation between percolation thresholds $p_c$ and coordination number $z$ in higher dimensions, we have carried out extensive Monte Carlo simulations for site and bond percolation on four-dimensional simple hypercubic lattice with extended neighborhoods up to 9th nearest neighbors. By employing an effective single-cluster growth method, we found precise estimates of the percolation thresholds for 18 total systems.

For site percolation, the asymptotic behavior was investigated by plotting both $z$ versus $1/p_c$ and $z$ versus $-1/\ln(1-p_c)$, and both figures for large $z$ tend to the theoretical value of $zp_c \sim 16 \eta_c \approx  2.086$ based upon $\eta_c = 0.1304(5)$ \cite{TorquatoJiao13}.  Our intercept value $c = 2.065$ implies $\eta_c = c/16 = 0.1291$ with an estimated error of $0.0007$, which is (nearly) within the combined standard deviations of the value 0.1304 of Ref.\ \cite{TorquatoJiao13}.

For bond percolation, the asymptotic value of $zp_c$ tends to the Bethe-lattice behavior of unity with the increase of the coordination number $z$. We accounted for the finite-$z$ corrections by considering Eqs.\ (\ref{eq:zpclnz}) and (\ref{eq:zpcx}), and our data shows consistency with the predicted behavior (\ref{eq:zpclnz}), but also a good fit with (\ref{eq:zpcx}) with $x = 3/4$. As the critical dimension for percolation is six, it will be interesting to further check the behavior of the bond thresholds in five dimensions.

\section{Acknowledgments}
The authors are grateful to the Advanced Analysis and Computation Center of CUMT for the award of CPU hours to accomplish this work. This work is supported by “the Fundamental Research Funds for the Central Universities” under Grant No. 2020ZDPYMS31.

\bibliographystyle{unsrt}
\bibliography{bibliography.bib}

\end{document}